\documentclass[
    twocolumn,
    prb,
    aps,
    citeautoscript,
    showpacs,
    footinbib,
    amsmath,
    amssymb,
    superscriptaddress,
    longbibliography
]{revtex4-2}

\usepackage[utf8]{inputenc}        % Input encoding
\usepackage[T1]{fontenc}           % Font encoding
\usepackage[english]{babel}        % Required for \selectlanguage in .bbl

\usepackage{amsmath, amssymb, amsfonts, bbold}

\usepackage{graphicx}
\usepackage{float}
\usepackage{dcolumn}
\usepackage{bm}
\usepackage{color}
\usepackage{array}
\usepackage{placeins}

\usepackage{natbib}                % For longbibliography (required)
\usepackage{hyperref}

\usepackage{physics}
\usepackage{braket}
\usepackage{miller}

\usepackage[normalem]{ulem}
\usepackage{tikz}

\newcommand{\beq}{\begin{equation}}
\newcommand{\eeq}{\end{equation}}
\newcommand{\beqa}{\begin{eqnarray}}
\newcommand{\eeqa}{\end{eqnarray}}

\begin{document}

% \title{Characterization of Yu-Shiba-Rusinov states in Gd nanostructures on superconducting Nb(110)}
\title{Rare-earth atoms on Nb(110) as a platform to engineer topological superconductivity}

\author{David Antognini Silva}
\thanks{These two authors contributed equally. The ordering of authors is alphabetical.}
\affiliation{Peter Gr\"{u}nberg Institut, Forschungszentrum J\"{u}lich and JARA, 52425 J\"{u}lich, Germany}
\affiliation{Institute for Theoretical Physics, RWTH Aachen University, 52074 Aachen, Germany}

\author{Yu Wang}
\thanks{These two authors contributed equally. The ordering of authors is alphabetical.}
% \email[]{yu.wang@physik.uni-wuerzburg.de}
\affiliation{Physikalisches Institut, Experimentelle Physik II,
Universit\"at W\"urzburg, Am Hubland, 97074 W\"urzburg, Germany}

\author{Nicolae Atodiresei}
\affiliation{Peter Gr\"{u}nberg Institut, Forschungszentrum J\"{u}lich and JARA, 52425 J\"{u}lich, Germany}

\author{Felix Friedrich}
\affiliation{Physikalisches Institut, Experimentelle Physik II,
Universit\"at W\"urzburg, Am Hubland, 97074 W\"urzburg, Germany}

\author{Stefan Bl\"ugel}
\affiliation{Peter Gr\"{u}nberg Institut, Forschungszentrum J\"{u}lich and JARA, 52425 J\"{u}lich, Germany}
\affiliation{Institute for Theoretical Physics, RWTH Aachen University, 52074 Aachen, Germany}

\author{Matthias Bode}
\affiliation{Physikalisches Institut, Experimentelle Physik II,
Universit\"at W\"urzburg, Am Hubland, 97074 W\"urzburg, Germany}
\affiliation{Wilhelm Conrad R\"ontgen-Center for Complex Material Systems (RCCM), Universit\"at W\"urzburg, Am Hubland, 97074 W\"urzburg, Germany}

\author{Philipp R\"{u}{\ss}mann}
% \email{p.ruessmann@fz-juelich.de}
\affiliation{Peter Gr\"{u}nberg Institut, Forschungszentrum J\"{u}lich and JARA, 52425 J\"{u}lich, Germany}
\affiliation{Institute of Theoretical Physics and Astrophysics, University of Würzburg, 97074 Würzburg, Germany}

% corresponding author:
\author{Artem Odobesko}
\email{artem.odobesko@physik.uni-wuerzburg.de}
\affiliation{Physikalisches Institut, Experimentelle Physik II,
Universit\"at W\"urzburg, Am Hubland, 97074 W\"urzburg, Germany}

\date{\today}

\begin{abstract}
Helical spin textures in one-dimensional magnetic chains on superconductors can enable topological superconductivity and host Majorana zero modes, independent of the presence of intrinsic spin–orbit coupling. Here, we show that gadolinium (Gd) adatoms, possessing large 4f magnetic moments when  placed on a Nb(110) surface, establish indirect exchange interactions mediated by  valence electrons, manifesting as Yu–Shiba–Rusinov states. By combining scanning tunneling microscopy and spectroscopy with density functional theory, we analyze the emergence of the Yu-Shiba-Rusinov states in single Gd atoms and Gd dimers and uncover the underlying magnetic interaction mechanisms, on the basis of which we predict by means of spin-dynamics simulations the formation of stable chiral Néel-type spin-spiral configurations in Gd chains. These findings highlight rare-earth magnets as a promising platform for precisely tuning spin-spiral ground states, a essential prerequisite for the realization of topological superconductivity.

    %Our study reveals how Gd adatoms and dimers on a superconducting Nb(110) surface induce Yu-Shiba-Rusinov (YSR) states, offering valuable insights into magnetic interactions of rare-earth atoms on superconducting surfaces. By engineering Gd dimers along the \hkl[1-10] and \hkl[001] directions, we uncover an indirect coupling between the Gd magnetic moments and the Nb substrate via their valence $d$ electrons, leading to significant alterations in the YSR spectrum around the dimers. We further demonstrate the possibility for N\'eel-type spin-spiral ground states in chains of Gd atoms on Nb(110). These findings highlight the potential of $4f$ elements like Gd as a promising platform for controlling a spin-spiral ground state, a crucial prerequisite for realizing a topological superconductor that can host Majorana zero modes. The combination of theoretical modeling based on density functional theory, atomistic spin-dynamics simulations and experimental techniques, including scanning tunneling microscopy and spectroscopy, provides a comprehensive understanding of the coupling mechanisms and their impact on the electronic properties of these systems and establishes rare-earth magnets on Nb as a promising platform in the field. 
\end{abstract}


\maketitle

\section{Introduction}

Yu-Shiba-Rusinov (YSR) states~\cite{Yu1965, Shiba1968, Rusinov1969}, which arise from the pair-breaking scattering of Cooper pairs off magnetic impurities in superconductors, have become a key focus in nanotechnology due to their role in forming topological 
superconductors (TSCs)~\cite{Balatsky2006}. These TSCs, formed by chains of magnetic atoms on superconducting surfaces, are predicted to host Majorana zero modes (MZMs), potentially enabling  fault-tolerant quantum computing via topological qubits~\cite{Braunecker2013, NadjPerge2013, Klinovaja2013a, Pientka2013}.
The interplay between magnetism and superconductivity is crucial for generating YSR states and achieving a topological phase.
Chains with a non-collinear spin configuration can support robust MZMs, as they inherently break time-reversal symmetry and introduce a synthetic spin-orbit coupling (SOC) effect, which is essential for stabilizing the topological phase ~\cite{Braunecker2013, Klinovaja2013, martin2012, Vazifeh2013, Schmid2022}.
In contrast, a ferromagnetic (FM) or antiferromagnetic (AFM) alignment typically keeps the system in a trivial, non-topological phase, unless strong SOC is present. Notably, in superconducting Nb(110)---a platform widely used for studying magnetic nanostructures on superconductors---SOC is virtually absent and does not influence its superconducting properties~\cite{Ruessmann2022}. Additionally, even if a topological phase can be achieved in FM/AFM chains with the aid of SOC, the resulting topological gap is often small, making the realisation of MZMs less robust to perturbations. Thus, precise control of the  magnetic ordering in these chains is essential for tuning the topological phase.

\begin{figure}[b]
	\includegraphics[width=0.7\linewidth]{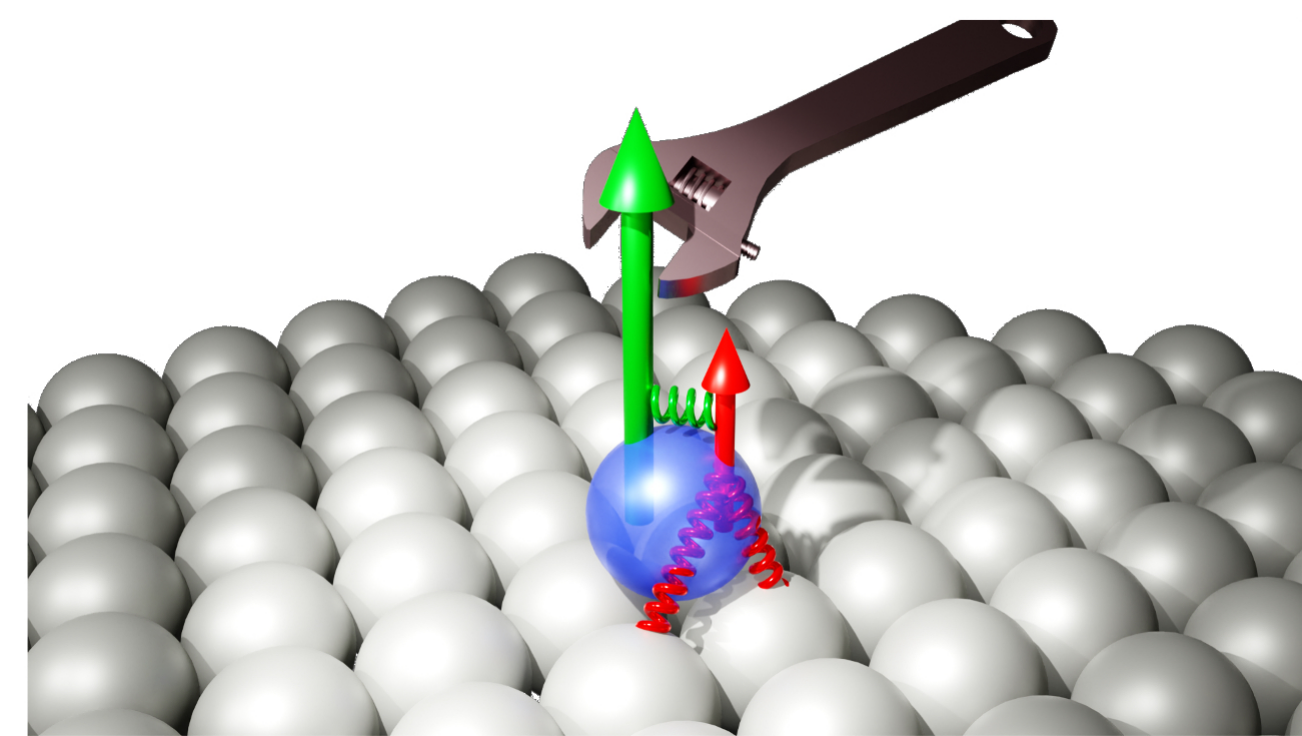}
	\caption{
 %Chains of rare-earth atoms as building blocks for topological superconductivity. \textbf{a} Illustration of a topologically trivial FM chain and a nontrivial helical chain hosting Majorana zero modes at the chain ends. \textbf{b} 
 Sketch of the indirect interaction of $4f$ (green arrow) and $5d$ (red arrow) moments in Gd to the Nb substrate. The torque wrench illustrates that the large $f$ moment of Gd is a big lever to control magnetism and thus topology in magnetic chains and nanostructures.}
	\label{sketch}
\end{figure}

In this context, rare-earth (RE) magnetic atoms on superconducting Nb(110) offer unique advantages for realizing spin-spiral chains. Unlike transition metal impurities, which have been extensively studied in the field~\cite{ruby2016, choi2018, Odobesko2020, Schneider2019, Menard2015, Kuester2021, Cornils2017, Beck2021, beck2023, friedrich2021, ruby2018, Nyari2021, nadj2014, ruby2017, kim2018, liebhaber2022, schneider2021, schneider2022precursors, Kamlapure2018, Nyari2023}, rare-earth atoms can possess very large magnetic moments due to their partially occupied $4f$ electron shells. These $4f$ electrons are highly localized around the rare-earth nucleus, meaning that the $4f$ moment couples to the superconducting substrate indirectly, mediated by the more delocalized valence $5d$ and $6sp$ electrons~\cite{Wienholdt2013, Pivetta2020}. This indirect coupling mechanism (sketched in Fig.~\ref{sketch}) introduces a degree of flexibility in how the magnetic moments interact with the superconductor, influencing the properties of the YSR states and, consequently, the topological characteristics of the system. So far, the study of rare-earth elements on superconductors has been limited to the early observation of YSR states at isolated Gd atoms on Nb(110)~\cite{Yazdani1997} and Bi(110)/Nb(110) heterostructures~\cite{Ding2021}. This highlights the need for a more comprehensive investigation of RE adatoms on superconductors.

Our work advances this frontier by demonstrating the potential of rare-earth atom nanostructures placed on superconducting surfaces as a platform for exploring topological superconductivity. %We focus our study on Gd single atom and dimers on Nb(110).
To establish how a $4f$ moment couples to a superconductor at the atomic scale, we begin by examining single Gd atoms and dimers on Nb(110) and use these elementary building blocks to gain fundamental insights that can guide the construction of more complex rare-earth chains. 
The electronic configuration of Gd provides a half-filled $4f$ shell with an additional occupied $5d$ electron, which maximizes the spin moment while simultaneously minimizing the orbital moment \cite{Schuh2012}. In the spirit of the Bruno model~\cite{Bruno1989}, this suggests a weak magnetocrystalline anisotropy~\cite{ColarietiTosti2003}, which can be advantageous in the realisation of a spin-spiral ground state in magnetic chains.
In fact, Gd compounds are known to stabilize non-collinear magnetic ground states due to frustrated exchange  by the RKKY interaction or  short-range superexchange, as it happens, for example, in  rare-earth intermetallics or Gd-based Weyl semimetals~\cite{Bouaziz2022, Hartl2022, Bouaziz2023, Hartl2024}.
Furthermore, the large $4f$ magnetic moments of Gd provide a strong lever for the potential engineering of magnetic fields on the magnetic ground state in chains of rare-earth atoms (see Fig.~\ref{sketch}\textbf{b}). This could allow for the design of tailored magnetic configurations, enabling control over the pitch of potentially arising spin spirals, which is a central quantity to realize the topological properties of helical chains on superconductors~\cite{Klinovaja2013a, Vazifeh2013}.

In this work we combine a detailed structural, magnetic and electronic analysis based on first-principles relaxed geometries with measured and simulated spectra of the arising YSR states inside the superconducting gap. Our work indicates the indirect coupling of the Gd $4f$ magnetic moments to the SC substrate via its valence $5d$ electron and shows how YSR states interact in dimers with a small separation between the magnetic adatoms. Finally we also explore the potential of Gd chains to form spin-spirals based on atomistic spin-dynamics simulations.

\section{Results}

\subsection{YSR states of Gd atoms and dimers}
\label{sec:STM}

For our scanning tunneling microscopy (STM) experiments, we deposited a small concentration of  Gd atoms on a clean Nb(110) surface at $4.2\,\mathrm{K}$ temperature. Using STM and STS, we examine the position and adsorption sites of the Gd atoms on the Nb substrate and probe the YSR states induced by the Gd isolated atom and dimers. Experimental details can be found in the ``Methods'' section and the Supplementary Material~\cite{supplement}.

\begin{figure*}
\centering
	\includegraphics[width=0.6\linewidth]{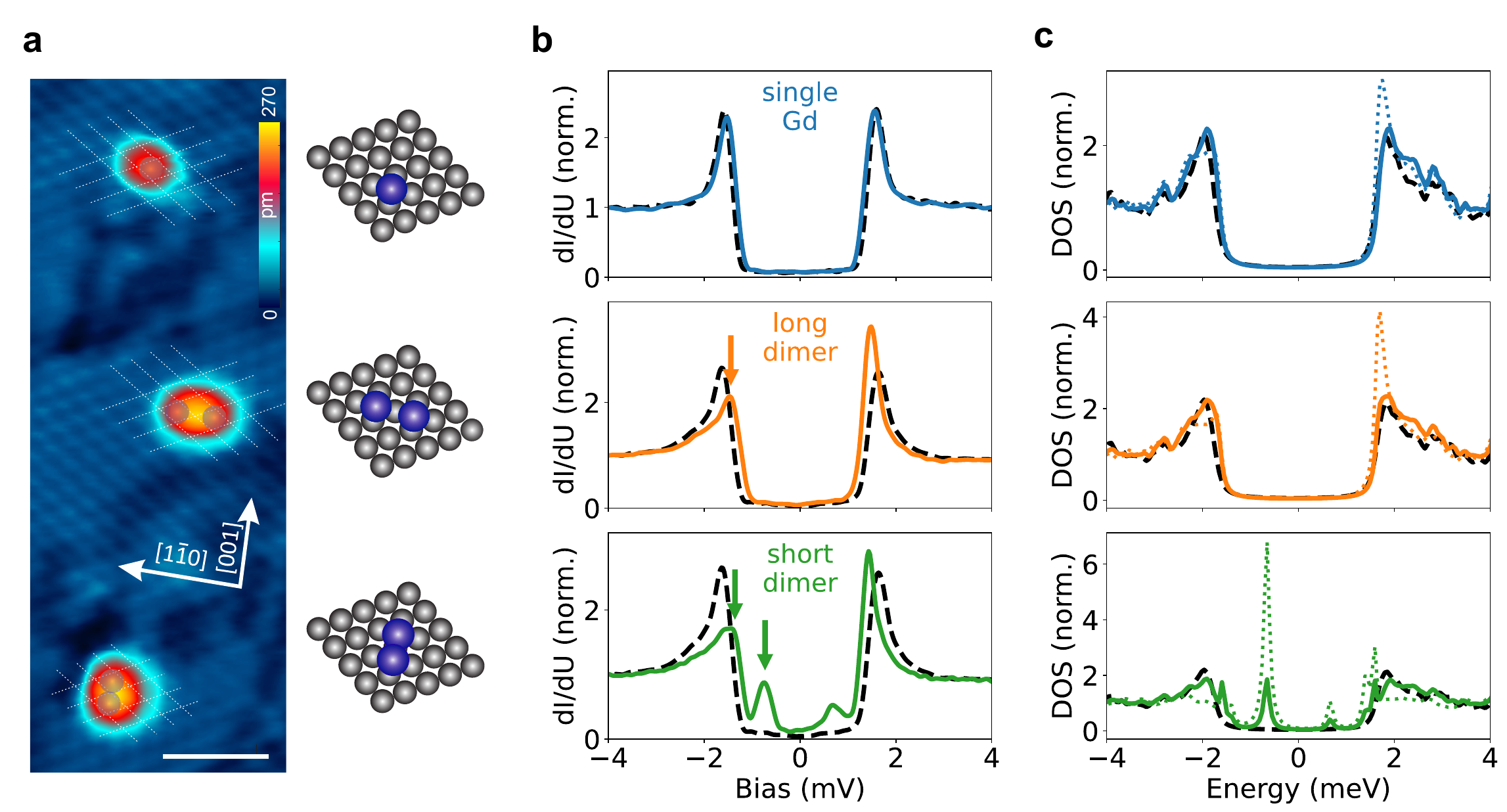}
	\caption{Characterization of YSR spectra of Gd atoms on a clean Nb(110) surface.
		\textbf{a} Atomic resolution of the \mbox{(110)} surface with self-assembled Gd structures (from top to bottom: single atom, long dimer, short dimer corresponding to the structural models on the right) recorded with a CO-functionalized tip. The white scale bar represents $2\,\mathrm{nm}$.
		\textbf{b} Deconvoluted tunneling spectra of (from top to bottom) the single Gd atom and the two Gd dimers (see Supplementary Material for raw data~\cite{supplement}). The black dashed line represents the spectrum on clean Nb(110). \textbf{c} Corresponding calculation results for YSR states. The black broken lines represent the calculated bare Nb DOS, while the full (dotted) coloured lines represent the calculated DOS integrated over the near impurity region (only the contribution of the Gd atom). The experimental and theoretical spectra are normalized by the $\dd I/\dd U$ (DOS) value averaged over $-5$ to $-4\,\mathrm{meV}$.}
	\label{dimers}
\end{figure*}

Figure~\ref{dimers}\textbf{a} presents a constant-current STM image of the Nb(110) surface with a single Gd atom and two  Gd-dimer configurations, captured using a CO-functionalized tip to enhance spatial resolution~\cite{friedrich2021}. The atomically resolved body-centered cubic (110) structure of Nb, with a lattice constant of $a = 3.32$ \AA\, is clearly visible. The Gd atoms, appearing as bright protrusions, show a much larger apparent size compared to the Nb unit cell. This enlargement likely arises from the large radius of the Gd $6s$ and $p$ orbitals, making it challenging to precisely determine the adsorption site. However, based on the surrounding crystalline structure, we infer that the single Gd atom occupies the fourfold hollow sites of the Nb(110) lattice. This conjecture is supported by our DFT calculations (see ``Methods'' section and the Supplementary Material~\cite{supplement} for computational details). We find that single Gd atoms bind strongly to the Nb(110) substrate ($4.5\,\mathrm{eV}$ binding energy), occupying the aforementioned hollow site with a distance of $2.4\,\mathrm{\AA}$ from the Nb(110) surface. 

In the middle and bottom part of Fig.~\ref{dimers}\textbf{a}, two self-assembled Gd dimers are visible, oriented along the \hkl[1-10] and \hkl[001] directions of the substrate, which we also refer to as ``long'' and ``short'' dimer, respectively, in the remainder of the manuscript. This is seen by the elongated shape along the dimer axis compared to the single Gd atom visible on the top of Fig.~\ref{dimers}\textbf{a}. Our DFT simulations reveal that the bonding of Gd to Nb is stronger than to Gd. We find that the binding energy per Gd atom of the short (long) dimer is $-3.9$~eV ($-4.4$~eV), and thus less than that of the single Gd adatom
%\sout{Our DFT simulations for Gd dimers reveal that the atoms generally attract each other . However, when absorbed on the Nb surface, the atoms are  forced to be located in distances following the Nb(110) lattice. Thus, compared to the gas-phase dimers, they experience either an added attraction or repulsion (depending on the dimer's direction) due to the shorter (longer) Nb--Nb distances along the \hkl[001] (\hkl[1-10]) directions on the surface compared to the preferred Gd--Gd distance in free-floating dimers}} 
(see Supplementary Material for details~\cite{supplement}). Thus, the dimers are metastable relative to the random distribution of Gd adatoms, but the strong binding to the substrate keeps the Gd atoms in place, allowing only minor in-plane relaxations. The Gd--Gd bondlength is $3.74\,\mathrm{\AA}$ ($4.14\,\mathrm{\AA}$) for the short (long) dimer, resulting in comparison to the ideal hollow-site positions of the Nb surface to a small in-plane outward (inward) relaxation of $0.21\,\mathrm{\AA}$ ($-0.28\,\mathrm{\AA}$), with negligible out-of-plane movement ($<0.05\,\mathrm{\AA}$). 

Figure ~\ref{dimers}\textbf{b} shows the deconvoluted tunneling spectra of YSR states measured on the  atomic configurations  presented in Fig.~\ref{dimers}\textbf{a} using a superconducting tip. The original spectra are provided in the Supplementary Material~\cite{supplement}. For a single Gd atom, no clear in-gap states are observed, and the spectrum is identical to that of clean Nb(110) (see blue and black dashed lines in Fig.~\ref{dimers}\textbf{b}). The absence of YSR states appears to contradict Yazdani \textit{et al.}~\cite{Yazdani1997}, which reported YSR signatures near the Nb coherence peaks.

For clarification, we deposited Gd on oxygen reconstructed Nb following the method in Ref.~\cite{Yazdani1997}. We identified two species: one without YSR states and another with YSR states near the gap edges (see Fig.~S3 of the Supplementary Material). Their apparent height suggests that the YSR-active species could be either unresolved Gd dimers or a single Gd atoms adsorbed at sites of altered hybridisation with the substrate due to the presence of oxygen, as reported in Ref.~\cite{Odobesko2020}
%\PRnote{@Artem Please decide how to write the following part about the comparison with Yazdani's work.}\AO{The apparent absence of YSR states at Gd seemingly disagrees with the previous report by Yazdani \textit{et al.}~\cite{Yazdani1997} where YSR signatures close to the Nb coherence peaks are observed. A comparison of Gd on an oxygen-reconstructed surface (Fig.~S3 of the Supplementary Material \cite{supplement}) \AOnote{, we believe that the data did not correspond to a single Gd atom. See more details in Supplementary materials.} showed no qualitative change in the spectra which rules out the possibility that the presence of oxygen at the surface~\cite{Odobesko2019}  drastically increases the degree of hybridization of the magnetic adatom with the substrate as it is the case for transition metal adatoms~\cite{Odobesko2020}.
%\PRnote{Please check this as possible explanations:}\PR{Instead, the dependence on the tip-sample distance~\cite{Yazdani1997} and other experimental parameters might prevent us from resolving the YSR states of isolated Gd adatoms.}}

We compare the tunneling spectra of two Gd dimers of different structural orientations. In the spectrum of the \hkl[1-10]-oriented long dimer, hich has a larger interatomic distance (orange line), we observe a slight shift in the $\dd I/\dd U$ signal near the coherence peaks, with an enhanced resonance at positive energy and a weaker one at negative energy-—indicative of YSR states. While hybridization of the YSR states in dimers typically leads to energy splitting \cite{Morr2003}, only a single pair of YSR features is clearly resolved  near the gap edges (orange arrow).
In contrast, the \hkl[001]-oriented short dimer (green line) exhibits an additional pair of in-gap peaks (green arrows). The stronger pair of resonances is located near the coherence peaks, similar to the \hkl[1-10]-oriented dimer, but a second, weaker pair emerges at a smaller bias voltage of $U = \pm(0.78 \pm 0.1)\,\mathrm{mV}$.

To better understand the YSR states observed in our STM experiments, we solve the Kohn-Sham Bogoliubov-de Gennes (KS-BdG) equations for magnetic Gd single atom and ferromagnetic dimers on Nb(110) (for details see ``Methods'' section)~\cite{jukkr2022, Ruessmann2022} and present the computed YSR spectra in Fig.~\ref{dimers}\textbf{c}.
Unlike the STM measurements, the KS-BdG simulations for a single Gd atom on Nb(110) are able to resolve YSR states at the edge of the superconducting coherence peak (dotted line), with an intensity asymmetry typical for YSR states. 
The very small induced moment of only $0.03\,\mu_B$ in the nearest Nb neighbor lets us conclude that the YSR states are not arising from the induced moment in Nb, but that the Cooper pairs are instead broken at the magnetic Gd impurities, as known from the full Gd overlayer~\cite{Park2024}. This interpretation is also in line with the coherence length of Nb that greatly exceeds the inter-atomic distances in Gd/Nb(110).
We suspect that the experimental broadening and the fact that the STM tip probes the local DOS in the vacuum above the sample might explain the apparent discrepancy between simulated and measured spectra for single Gd atom. This conjecture is supported by the fact that, when averaging the signal over the near impurity region (i.e., including the Gd atom and contributions from surrounding Nb and vacuum sites), the calculated DOS shows very little difference compared to the bare Nb. The resulting data for the near impurity region (full line in Fig.~\ref{dimers}\textbf{c}) then reproduce the experimental data (Fig.~\ref{dimers}\textbf{b}) very well.

In the long dimer configuration, the calculated YSR peak asymmetry grows, and the YSR states shift slightly to lower binding energies. For the short dimer, a prominent YSR state appears at $\pm0.65\,\mathrm{meV}$ within the superconducting gap, along with additional, smaller YSR peaks splitting off from the coherence peaks. The location of the inner YSR peak agrees very well with the measured peak position %at $\pm(0.78 \pm 0.1)\,\mathrm{meV}$. 
Accounting for thermal broadening, the observed position and intensity asymmetries of the calculated YSR states agree well with the experimental data.

\subsection{Origin of YSR states and their interaction}

A deeper understanding of the YSR states can be obtained from the orbital decomposition of the superconducting DOS of the short Gd dimer (Fig.~\ref{fig:fig3}\textbf{a}), which exhibits the clearest YSR spectrum with pronounced in-gap states (see Supplementary Material for single Gd atom and long dimer~\cite{supplement}).
The orbital resolution reveals that the observed YSR features originate from three distinct pairs of YSR states ($\varepsilon_1$, $\varepsilon_2$, $\varepsilon_3$) of different orbital characters: %\SBc{Of which orbital characters of Gd do you talk here , of the Gd majority states, minority states, or of both because there spin solitting is so tiny ??, this is inportant }: 
$\varepsilon_1$ has a strong contribution from $d_{x^2-y^2}$ orbitals of Gd and $d_{z^2}$ orbitals of nearby Nb atoms (shown in the Supplementary Material~\cite{supplement}), $\varepsilon_2$ predominantly comes from $d_{xz}$ orbitals in Gd, and $\varepsilon_3$ has a strong mixing with $d_{xy}$ and $d_{yz}$ orbitals of Gd and Nb neighbors, respectively. These observations support our hypothesis that the Gd $f$-shell alone does not induce YSR states due to its lack of direct coupling to the superconducting host. Instead, the localized $4f$ states induce an intra-atomic ferromagnetic spin-polarization of the Gd valence $5d$ electrons, which, in turn, mediate the interaction with the superconducting states of Nb.

\begin{figure}
	\centering
	\includegraphics[width=\linewidth]{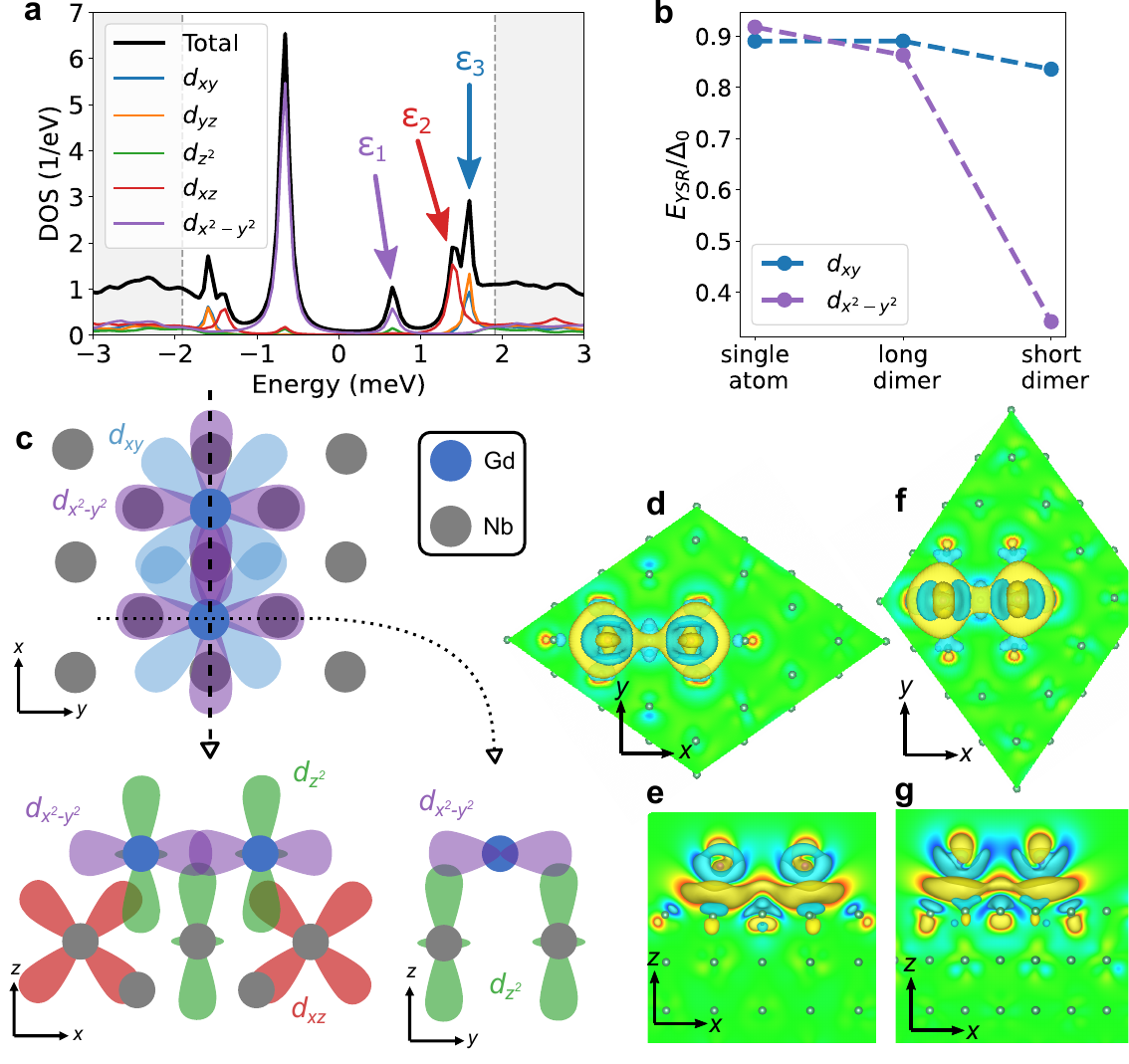}
	\caption{Orbital decomposition and interaction of YSR states. \textbf{a} Orbital-resolved superconducting DOS of short Gd dimer. The $x$ and $z$ directions are defined as the connecting vector of the Gd atoms and the Nb surface normal, respectively. The grey area marks the edges of the Nb(110) host's superconducting gap identified from the main coherence peak of Nb. \textbf{b} Changes in the position of the YSR states for single adatom and the different Gd dimers. \textbf{c} Sketch (top view) of the orbital hybridization in the short Gd dimer with the Nb substrate. \textbf{d-g} Visualization of the charge redistribution and transfer in the long (\textbf{d, e}) and short (\textbf{f, g}) Gd dimers compared to isolated Gd adatoms on Nb(110). The yellow/ red color represents the charge density accumulation and the light blue / blue color indicates a charge density depletion upon bonding.} %\SBc{Wouldn't the difference between a dimer on Nb and two adatoms on Nb be interesting? Is the color coder in terms of chanrge density changes teh same for all 4 figure d-g or are they scaled differntly ? It would be better to add a color code bar or put it at least in the supplement and refer to it. }}
	\label{fig:fig3}
\end{figure}

Figure~\ref{fig:fig3}\textbf{b} presents the energy positions of the YSR peaks with in-plane orbital character for different Gd dimers. We focus on comparing YSR states originating from the $d_{xy}$ and $d_{x^2-y^2}$ orbitals, as they maintain a consistent relative orientation with respect to both the Nb substrate and the dimer direction. In contrast, YSR states associated with orbitals such as $d_{xz}$ (where x is along the dimer direction) may exhibit energy variations due to differences in relative orientation with the Nb lattice. 

By restricting our comparison to the $d_{xy}$ and $d_{x^2-y^2}$ orbitals, we observe only minor energy shifts between the single Gd atom and the long Gd dimer. However, the short Gd dimer exhibits a significant shift, which can likely be attributed to the orbital pointing along the dimer direction, as this orientation modifies the hybridisation strength between Gd atoms and the Nb substrate, ultimately influencing the YSR energy position. This refined analysis provides a more accurate interpretation of the orbital-dependent YSR energy shifts in Gd dimers on Nb(110), though additional factors may also contribute.

%Fig.~\ref{fig:fig3}\textbf{b} reports the energy position of the YSR peaks \DA{\sout{for} of in-plane orbital character for} the different Gd systems. \DA{[Note: We decide to only compare YSR formed by those orbitals because of the Nb anisotropy and the different dimer orientation. Example: the YSR state formed by the $d_{xz}$ orbital (x being along the dimer direction), even if present in both short and long dimers, has a different orientation compared to the Nb substrate. Therefore, the changes in energy we observe might only come from their different relative orientation with the substrate. So, only the $d_{xy}$ and $d_{x^2-y^2}$ can be directly compared from one system to the other, because they keep the same relative orientation with both the Nb substrate and the dimer direction.]} While we only observe minor changes in the YSR energies between the single Gd atom and the long dimer, the short Gd dimer exhibits a significant shift, which we associate with the substantial change in the hybridization between \SB{the Gd atoms in the short dimer and the Gd atoms in the dimer with} the Nb substrate. \DA{[Note: With the new labeling and the proper orbital comparison, I dont think this is true anymore. The largest shift is observed for the orbital that points directly along the dimer direction.]}

Qualitatively, the energy shift of YSR states can be understood using a simple model for bound states~\cite{Balatsky2006, Saunderson2022}:
\begin{equation}
    \varepsilon_i = \pm \Delta_0\frac{1-\tilde\alpha_i^2}{1+\tilde\alpha_i^2},
    \label{eq:YSR}
\end{equation}

where $\Delta_0$ is the superconducting gap of the host, and $\tilde\alpha_i=\pi \mathcal{N}_{0,i} J_i S_i$ describes the effective coupling, which depends on the normal-state electron density $\mathcal{N}_0$, the exchange interaction $J$, and the impurity spin $S$. Since these parameters are orbital-dependent (indicated by the subscript $i$), variations in YSR energy shifts can arise from differences in hybridization strength.

%where $\Delta_0$ is the size of the host's superconducting gap, and $\tilde\alpha_i=\pi \mathcal{N}_{0,i} J_i S_i$ is the effective coupling parameter that connects the host's normal state electron density $\mathcal{N}_0$ via a coupling term $J$ of the impurity's spin $S$ to the substrate. Note that these are orbital-dependent quantities as indicated by the subscript $i$ in Eq.~\eqref{eq:YSR}~\cite{Saunderson2022}.The largest change is observed for the in-plane-oriented \DA{$d_{x^2-y^2}$} orbitals that hybridize with Nb $d_{z^2}$ orbitals.

The most pronounced change is observed for the $d_{x^2-y^2}$ orbitals, which hybridize with Nb $d_{z^2}$ orbitals. This hybridization difference is already evident in the normal-state charge transfer, as shown in Fig.\ref{fig:fig3}\textbf{d-g}, where charge accumulation and depletion patterns around Gd atoms differ between the short and long dimers (see also Supplementary Material\cite{supplement}). At the Gd sites, charge density accumulates in the $d_{xy}$ and $d_{x^2-y^2}$ in-plane orbitals and the $d_{z^2}$ and $d_{yz}$ out-of-plane orbitals, while a notable depletion occurs in the $d_{xz}$ orbital, as indicated by the larger blue area in Fig.~\ref{fig:fig3}\textbf{e, g}. Importantly, these charge distribution differences between the short and long dimers suggest a distance-dependent modification of hybridization.

%This change in the hybridization is already seen in the normal state charge transfer, shown in Fig.~\ref{fig:fig3}\textbf{d-g} around the Gd atoms for the long and short dimers (see also the Supplementary Material~\cite{supplement}).At the Gd sites, we observe a charge density accumulation in the $d_{xy}$ and $d_{x^2-y^2}$ in-plane as well as $d_{z^2}$ and $d_{yz}$ out-of-plane orbitals while there is a clear charge density depletion in the $d_{xz}$ out-of-plane orbitals, as indicated by the larger blue area in Fig.~\ref{fig:fig3}\textbf{e, g}. Importantly, this is modified significantly from the short to the long dimer (cf.\ Figs.~\textbf{e, g}). \SBc{Honestly the charge density differences due to bonding of the long and short dimer is rather small but the change of $\epsilon_1$ is very large, is your argument convincing?} Thus, the charge density accumulation and depletion due to hybridization with the Nb(110) substrate in different $d$-orbitals at the Gd site crucially depends on the Gd–Gd distance explaining the striking difference between the spectra of the short and long dimers.

We interpret the shift of the YSR states as a consequence of changes in hybridization with the substrate, rather than a bonding/antibonding splitting of YSR states originating from neighboring Gd atoms. While previous theoretical models of interacting YSR states suggest a bonding/antibonding splitting mechanism\cite{Morr2003}, this interpretation is not supported in our case, as no clear even/odd features are observed in the spatial mapping of YSR wavefunctions around the Gd dimers (cf. Fig.~S4 of the Supplementary Material~\cite{supplement}). The absence of such symmetry-based signatures, which are typically linked to the spatial structure of atomic orbitals that mediate the exchange interaction and give rise to magnetic scattering~\cite{ruby2016, kuster2022non}, suggests a fundamental difference between the $4f$ Gd adatom and $3d$ transition metal impurities on Nb(110)~\cite{friedrich2021, beck2023, choi2017}.

%Importantly we interpret the shift of the YSR states to a change in the hybridization with the substrate and not as a result of a bonding/antibonding splitting of the YSR states stemming from nearby Gd atoms. \SBc{I am not sure this is convincing ,  I see a different bonding in $xz$.} This seemingly contradicts earlier theoretical modeling of interacting YSR states~\cite{Morr2003}.  This interpretation is supported by the absence of discernible even/odd features in the spatial mapping of YSR wavefunctions around the Gd dimers (cf.\ Fig.~\DA{S1} of the Supplementary Material \cite{supplement}) preventing the identification of the atomic orbital shape responsible for the potential magnetic scattering channel~\cite{ruby2016, kuster2022non}. This highlights a clear difference of the $4f$ Gd adatom compared to various $3d$ transition metal atoms on Nb(110)~\cite{friedrich2021,beck2023,choi2017}.

This interpretation is further supported by our theoretical calculations, which also fail to resolve any clear even/odd symmetry for the YSR states. The absence of such characteristics in $f$-element Gd highlights the complexity of its indirect interaction with the substrate, which occurs through a combination of $5d6s$ valence orbitals. Overall, these observations suggest that the interaction between the valence $d$ electrons of Gd and Nb atoms is responsible for the observed YSR states. The distance-dependent interaction with the substrate in the long and short dimers ultimately drives the energy shifts of the YSR peaks, though additional effects—such as subtle modifications in exchange interactions and local electronic correlations—may also contribute.

%These results \SBc{Which results , you mean your interpretation?} are consistent with our theoretical calculations where no clear even/odd symmetry for the different YSR states can be resolved. The absence of these characteristics for the $f$-element Gd further underscores the complexity of the indirect interaction with the substrate through a combination of $5d6s$ orbitals. Overall, these observations suggest that the interaction between valence $d$ electrons of Gd and Nb atoms is responsible for the observed YSR states. The interaction with the substrate, furthermore, depends on the Gd--Gd distance in the long or short dimer inducing, ultimately,  the changes in the energy of the YSR peaks.

\subsection{Towards topological superconductivity in rare-earth chains}

\begin{figure}
    \centering
    \includegraphics[width=0.9\linewidth]{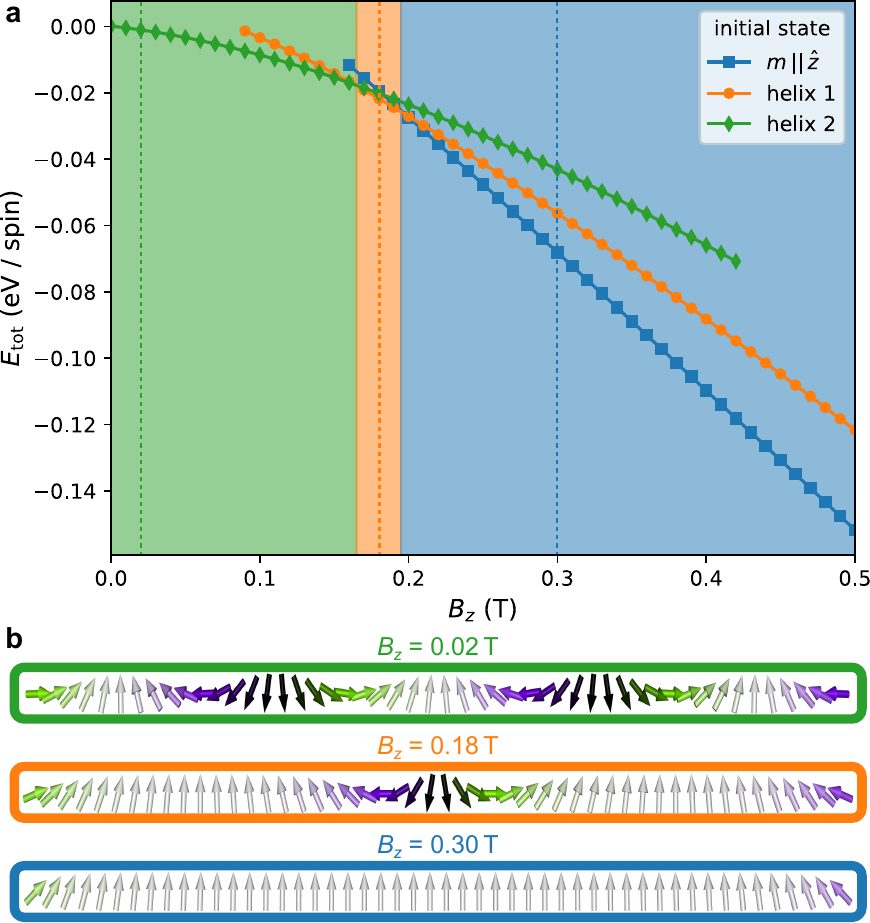}
    \caption{Energy of a linear chain of 50 Gd atoms on Nb(110) along the \hkl[1-10] direction of the long dimer. \textbf{a} Phase space of the magnetic ground state as function of the external magnetic field $B_z$. 
    The background color highlights the state with the lowest energy (green: $2.5$ rotations, orange: $1.5$ rotations, blue: no full rotation = FM state). Below $B_z^c=0.2\,\mathrm{T}$ a N\'eel-type spin spiral ground state is found.
    \textbf{b} Three representative spin configurations for the two helical and the FM ground state that stabilize at different external magnetic fields.}
    \label{fig:spirit}
\end{figure}

After characterizing the existence of YSR states in Gd dimers, we finally explore the possibility of a spin spiral ground state in Gd chains on superconducting Nb(110), which is a necessity to obtain a topologically nontrivial phase. Our DFT study reveals that the ferromagnetic (FM) state is preferable over the antiferromagnetic (AFM) for both long and short Gd dimers along the \hkl[1-10] and \hkl[001] directions, respectively. To gain further insights, we compute the exchange interactions (see ``Methods'' section and Supplementary Material~\cite{supplement} for details) using the method of infinitesimal rotations~\cite{Liechtenstein1987, Ebert2009} summarized in  Tab.~\ref{tab:Jij}. This allows to map the quantum mechanical DFT simulations onto the  extended Heisenberg Hamiltonian describing the interaction of classical spins $\mathbf{S}_i =\mathbf{M}_i/\mu_i$ ($\mu_i = |\mathbf{M}_i|$) sitting at lattice sites $i$ 
\begin{eqnarray}
    H = &-&\sum_{\braket{ij}} J_{ij} \left( \mathbf{S}_i \cdot \mathbf{S}_j \right) 
    % \nonumber\\
         -  \sum_{\braket{ij}} \mathbf{D}_{ij} \cdot \left( \mathbf{S}_i \times \mathbf{S}_j \right) 
    %\nonumber\\
        - \sum_{i} \mu_{i} \mathbf{B}\cdot \mathbf{S}_{i} \ 
    \label{eq:Heisenberg}
\end{eqnarray} 
Here, the first and second terms are the isotropic ($J_{ij}$) and chiral exchange interactions ($\mathbf{D}_{ij}$, also known as Dzyaloshinskii-Moriya interaction), and the last term describes the Zeeman energy due to an external magnetic field $\mathbf{B}$. We neglect the uniaxial magnetic anisotropy $K$ in the Gd chains that was found to be weak ($K <0.1\,\mathrm{meV}$ per Gd atom) and study chains of 50 Gd atoms using the computed nearest neighbor exchange interactions (indicated in \eqref{eq:Heisenberg} by $\braket{ij}$) from either short or long dimers as a toy model to explore the potential for realizing a spin-spiral magnetic ground state in chains of Gd atoms on Nb(110). Using a series of spin-dynamics calculations based on the Landau-Lifshitz-Gilbert (LLG) equations,
% ~\cite{LandauLifshitz, Gilbert},% comment out to get down to 70 references
we study the magnetic ground state in chains along either the \hkl[1-10] or \hkl[001] directions. For chains along the \hkl[001] (short) direction, we find a FM ground state even at vanishing magnetic fields due to the strong isotropic exchange interaction. However, for the \hkl[1-10] (long) direction, we find a homogeneous homochiral spin spiral ground state with a counterclockwise rotational sense (shown in Fig.~\ref{fig:spirit}),  which we can trace back to the sizable $D/J$ ratio ($D=|\mathbf{D}_{ij}|$), a ${D}_{ij}>0$ and a sufficiently small magnetic anisotropy ($D^2/J > K$), giving an estimated canting angle between two neighboring spins of $\theta=\arctan(D/J)\approx20^\circ$. With increasing the external field, the homogeneous spins spiral becomes inhomogeneous, is deformed into a large and small FM deomain separated by chiral domain walls until for larger fields $B_z \ge B_z^c=0.20\,\mathrm{T}$ the field saturated FM state without a full rotation of the moments becomes the ground state solution.

\begin{table}[]
    \centering
    \caption{Nearest neigbhor isotropic exchange ($J_{ij}$) and Dzyaloshinskii-Moriya interaction ($\mathbf{D}_{ij}$) for short and long Gd dimers located at site $i$ and $j$. The third and fourth columns report the $D/J$ ratio and the canting angle $\theta=\arctan(D/J)$. Note that, with the definition of that the $x$-direction follows the connecting vector of the two Gd atoms ($\mathbf{R}_{ij}/|\mathbf{R}_{ij}| = \mathbf{\hat{x}}$), the $\mathbf{D}_{ij}$ vector for both short and long dimer points in $y$-direction $\mathbf{D}_{ij}=(0, -D_{ij}, 0)^T \propto \mathbf{R}_{ij} \times \mathbf{\hat{z}}$ in accordance to Moriya's rule taking into account that the surface normal $\mathbf{\hat{z}}$ is the direction of the symmetry breaking.}
    \begin{tabular}{c|c|c|c|c}
                & $J_{ij}$ (meV) & ${D}_{ij}$ (meV) & $D/J$   & $\theta$ ($^\circ$) \\\hline
    short dimer & $30.42$        & $0.036$          & $0.001$ & $0.07$ \\
    long dimer  &  $1.18$        & $0.41$           & $0.35$  & $19.1$
    \end{tabular}
    \label{tab:Jij}
\end{table}

\begin{figure}
    \centering
    \includegraphics[width=\linewidth]{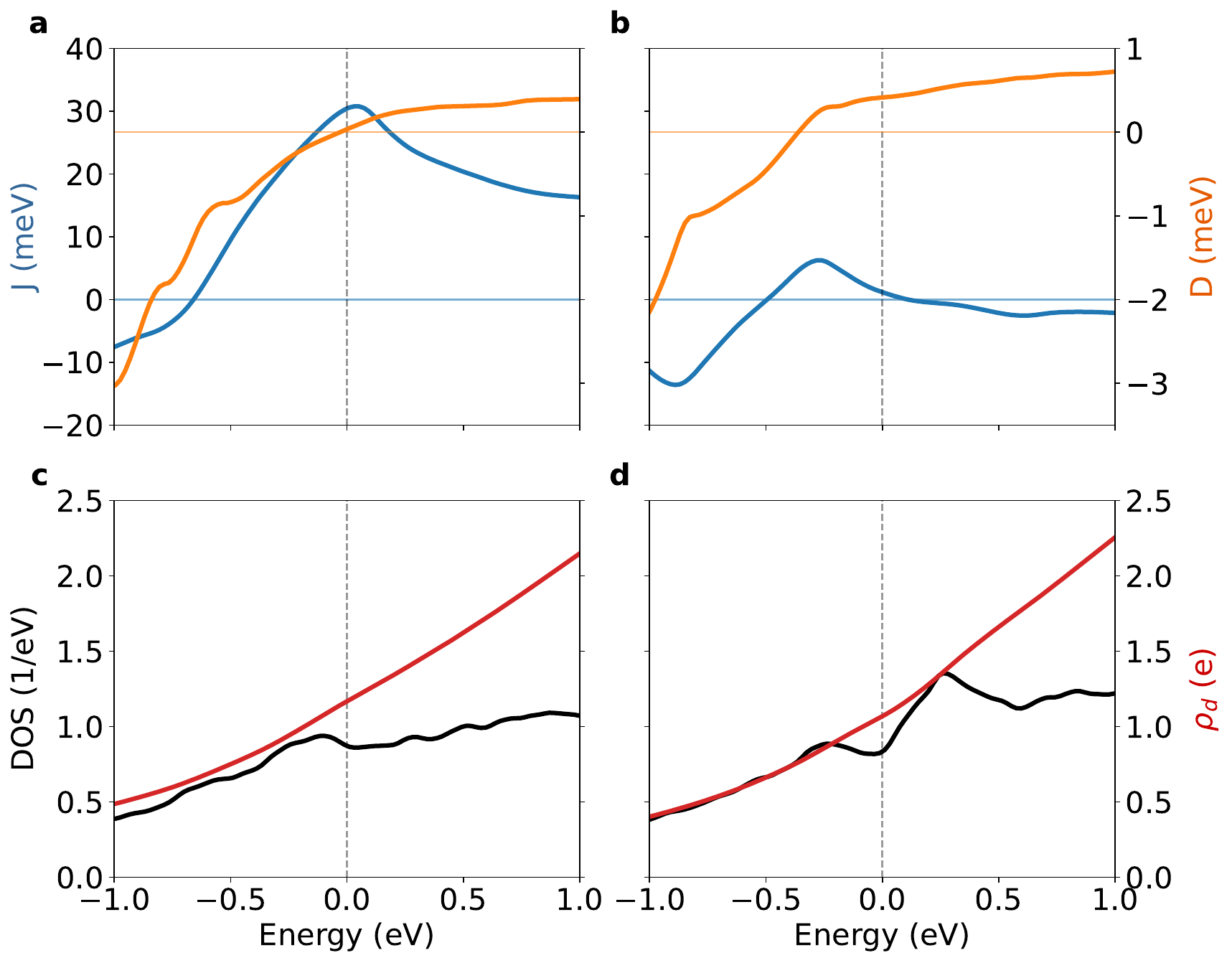}
    \caption{Nearest neighbor exchange ($J_{ij}$, blue) and Dzyaloshinskii-Moriya interaction ($D_{ij}=|\mathbf{D}_{ij}|$, orange) for \textbf{a} short and \textbf{b} long dimer as a function of a shifted Fermi energy. \textbf{c, d} Density of states (black) of the $d$ orbitals in the short and long dimers, respectively, together with the associated $d$ occupation (red) as function of a shifted Fermi energy.}
    \label{fig:Jij}
\end{figure}

The strengths of $J_{ij}$ and $\mathbf{D}_{ij}$ depend on the electronic configuration of the magnetic atom and there, crucially, on the filling of the Gd valence $d$-orbital. Figure~\ref{fig:Jij} shows the dependence of the exchange coupling parameters and the occupation of the Gd $d$-orbitals for both short and long dimers as a function of the Fermi level (see Supplementary Material for details~\cite{supplement}). This could be controlled in an experimental setup by either local electrical gating or by absorption of additional overlayers or molecules as done, for example, via the skyhook effect changing the charge state and thus the anisotropy of $4f$ metals on a heavy metal substrate~\cite{Herman2022}. On the one hand, Fig.~\ref{fig:Jij}\textbf{a} shows that for the short dimer, achieving $D/J$ ratio noticeably different from zero requires an unrealistically large shift of the Fermi level by $\approx-0.6\,\mathrm{eV}$ (where $J(E)$ shows a sign change and therefore crosses zero) and a corresponding large change in the $d$-occupation of $\approx0.4\,\mathrm{e}$ per Gd atom. On the other hand, the long dimer shows a much higher tunability where, already with small Fermi level shifts of $E\approx \pm50\,\mathrm{meV}$, the $D/J$ ratio can be approximately doubled or halved compared to the original Fermi level. This would allow to control the pitch of the spin spiral and the size of the critical magnetic field where the FM ground state becomes the most stable solution (see Supplementary Material~\cite{supplement}). The possibility to control the topological state of magnetic chains was previously only demonstrated theoretically in a hypothetical scenario for $3d$ adatom chains on Nb(110)~\cite{Nyari2023,Laszloffy2023}. Importantly, our findings demonstrate the potential to engineer the spin spiral ground state---and thus the topological nature---of Gd atom chains on Nb(110) using external electric and magnetic fields.

\section{Discussion}

In our combined theoretical and experimental study, we investigate individual Gd atoms and dimers on superconducting Nb(110). Despite the strong magnetic moment stemming from a half-filled $4f$-shell, an isolated Gd atom exhibits only weak YSR signatures within the superconducting gap, which become more pronounced in Gd dimers oriented along the \hkl[1-10] and \hkl[001] directions. This behaviour is attributed to the indirect interaction mediated by the valence $d$ orbitals, which couple the $4f$ moment to the superconducting substrate. Notably, only Gd dimers at the shortest possible separation (\hkl[001]) induce substantial changes in the YSR spectrum, where compressive strain from adsorption on Nb(110) significantly alters hybridization with the substrate.

Our investigations of the magnetic ground states in Gd atom chains on Nb(110) demonstrates that controlled atomic manipulation enables the formation of spin spirals when chains are constructed along the \hkl[1-10] direction. Moreover, the ability to tune exchange interactions, particularly the $D/J$ ratio via local gating, provides a means to control the spin-spiral pitch and, consequently, the regime of topological superconductivity in rare-earth-based magnetic chains. The large $4f$ moment of Gd, which interacts with the substrate only indirectly via its $5d$ electrons, further allows dynamic control of magnetic order, enabling transitions from a spin-spiral to a ferromagnetic ground state at small external fields. Crucially, these field strengths remain below the critical field of superconducting Nb films~\cite{Zaytseva2020}, offering a viable route to tunable Majorana-based platforms.

In conclusion, our study  highlights the distinct behaviour of rare-earth $f$-elements compared to conventional transition-metal impurities on Nb(110). The indirect coupling of RE atoms via intra-atomic spin-polarized valence $d$ electrons provides a novel approach to engineering tunable platforms for topological superconductivity. These findings advance our understanding of the interplay between magnetism and superconductivity and pave the way for the realization of controllable Majorana-based quantum computing architectures.

\section{Methods}

\subsection{Density functional theory simulations} \label{ssec:DTF}
 
Different types of density functional theory (DFT)
% ~\cite{T1, T5} % comment out to get down to 70 references
simulations for Gd nanostructures on Nb(110) are performed for this study using two different codes: (i) Structural relaxations have been carried out with the wave function code {\tt VASP}~\cite{T3,T4}, and (ii) the properties of the superconducting state are evaluated using the Kohn-Sham Bogoliubov-de Gennes (KS-BdG) methodology implemented in the relativistic full-potential DFT code {\tt JuKKR}~\cite{Ruessmann2022, jukkr2022}. 
The LDA+U~\cite{T7,Ebert2003} method is used for the proper treatment of  Gd~\cite{Kurz2002} with $U^\mathrm{eff} = U - J$ of $7$\,eV and 
the PBE exchange-correlation energy functional~\cite{T6} was employed. The mapping onto the extended Heisenberg Hamiltonian is done using the method of infinitesimal rotations~\cite{Liechtenstein1987, Ebert2009} and the ground state of the spin chains discussed in the main text are found with atomistic spin-dynamics calculations based on the Landau-Lifshitz-Gilbert (LLG) equation
% ~\cite{LandauLifshitz, Gilbert}% comment out to get down to 70 references
as implemented in the {\tt Spirit} code~\cite{spirit-paper}. Further computational details can be found in the Supplementary Material~\cite{supplement}.

\subsection{Experimental methods}

All measurements are conducted using a custom-made low-temperature STM at 1.4\,K. The clean Nb\mbox{(110)} surface is prepared through sputtering with Ar ions and a series of high-temperature annealing cycles~\cite{Odobesko2019}. Gd atoms are deposited \textit{in-situ} onto the cold Nb\mbox{(110)} substrate at a temperature of 4.2\,K. Spectroscopic measurements are conducted using a lock-in technique with a modulation voltage of 0.1\,mV at a modulation frequency of 890\,Hz. In order to enhance the energy resolution of the spectroscopic measurements, tungsten tips are functionalized 
with a Nb cluster on the tip apex (see Ref.~\onlinecite{Odobesko2020} for details of the experimental procedure). 
Further details can be found in the Supplementary Material~\cite{supplement}.

\begin{acknowledgments}
We thank Hao Wang for providing the initial value for the out-of-plane relaxated distance of single Gd from the Nb surface.
The work was supported by the Deutsche Forschungsgemeinschaft (DFG, German Research Foundation) through SFB 1170 (project C02, project-id 258499086), SFB 1238 (project C01) and under Germany's Excellence Strategy through W{\"u}rzburg-Dresden Cluster of Excellence on Complexity and Topology in Quantum Matter -- ct.qmat (EXC 2147, project-id 390858490) and the Cluster of Excellence Matter and Light for Quantum Computing -- ML4Q (EXC 2004/1, project-id 390534769). Furthermore, P.R.\ acknowledges the financial support by the Bavarian Ministry of Economic Affairs, Regional Development and Energy within Bavaria’s High-Tech Agenda Project ``Bausteine für das Quantencomputing auf Basis topologischer Materialien mit experimentellen und theoretischen Ansätzen''.
%{\sout{We also gratefully acknowledge the computing time granted through JARA on the supercomputer JURECA~\cite{jureca} at Forschungszentrum Jülich (projects ``superint'' and ``\PRnote{@Nico add your project id here}'')}}
The authors gratefully acknowledge the computing time granted by the
JARA Vergabegremium and provided on the JARA Partition part of the
supercomputer JURECA at Forschungszentrum Jülich.
\end{acknowledgments}

\section*{Code and Data availability}

The full provenance of the KS-BdG calculations, given by the connection of our {\tt JuKKR} and {\tt Spirit} codes to the AiiDA framework~\cite{aiida, aiida-kkr-paper, aiida-spirit}, which includes the values of all numerical cutoffs used in our simulations, is exported as a publicly accessible database in the Materials Cloud Archive~\cite{doi-dataset}. This dataset also contains the raw experimental data presented in this work.

\bibliography{bibliography.bib}

\end{document}